\documentclass[aps,prb,twocolumn,preprintnumbers,superscriptaddress,amsmath]{revtex4}
\usepackage{graphicx}
\usepackage{dcolumn}
\usepackage{bm}
\usepackage{soul}
\usepackage{color}
\usepackage{xcolor}
\usepackage{amsmath}
\usepackage{amssymb}
\usepackage{amsmath}
\usepackage{float}
\usepackage{natmove}
\usepackage{multirow}
\usepackage{setspace}
\usepackage{array}
\usepackage{booktabs}
\usepackage{rotating}
\usepackage{ulem}
\usepackage[colorlinks=true,linkcolor=blue,citecolor=blue,urlcolor=blue]{hyperref}
\usepackage{physics}
\usepackage{mathrsfs}
\definecolor{Grey}{rgb}{0.50,0.50,0.50}
\definecolor{Blu}{rgb}{0.00,0.00,1.00}
\definecolor{Red}{rgb}{1.00,0.00,0.00}
\definecolor{Green}{rgb}{0.00,0.60,0.00}
\definecolor{Magenta}{rgb}{0.60,0.00,0.60}
\definecolor{BluBondi}{rgb}{0.00,0.58,0.71}
\definecolor{Orange}{rgb}{0.95,0.46,0.17}

\newcommand{\editor}[2]{%
  \expandafter\newcommand\csname #1note\endcsname[1]{%
    \textcolor{#2}{(\textbf{#1:} \it ##1)}}%
  \expandafter\newcommand\csname #1\endcsname[1]{%
    \textcolor{#2}{##1}}%
  \expandafter\newcommand\csname #1cancel\endcsname[1]{%
    \textcolor{#2}{\sout{##1}}}%
  \expandafter\newcommand\csname #1change\endcsname[2]{%
    \textcolor{#2}{\sout{##1} ##2}}%
  \newenvironment{#1text}{\color{#2}}{\color{black}}
}

\editor{MB}{orange}
\editor{EM}{BluBondi}
\editor{AR}{magenta}
\editor{MJC}{Green}
\editor{AF}{BluBondi}
\editor{DV}{Blu}

\newcommand{\suppinfo}{Supplemental Material~\cite{supp-info}}
\newcommand{\drop}[1]{}

\newcommand{\kk}{\mathbf{k}}
\newcommand{\rr}{\mathbf{r}}

\begin{document}
\title{Excitonic effects in graphene-like C$_{3}$N}

\author{Miki Bonacci}
\email[corresponding author:]{miki.bonacci@unimore.it}
\affiliation{Dipartimento di Scienze Fisiche, Informatiche e Matematiche, Universit$\grave{a}$ di Modena e Reggio Emilia, I-41125 Modena, Italy}
\affiliation{Centro S3, CNR-Istituto Nanoscienze, I-41125 Modena, Italy}
\author{Matteo Zanfrognini}
\affiliation{Dipartimento di Scienze Fisiche, Informatiche e Matematiche, Universit$\grave{a}$ di Modena e Reggio Emilia, I-41125 Modena, Italy}
\affiliation{Centro S3, CNR-Istituto Nanoscienze, I-41125 Modena, Italy}
\author{Elisa Molinari}
\affiliation{Dipartimento di Scienze Fisiche, Informatiche e Matematiche, Universit$\grave{a}$ di Modena e Reggio Emilia, I-41125 Modena, Italy}
\affiliation{Centro S3, CNR-Istituto Nanoscienze, I-41125 Modena, Italy}
\author{Alice Ruini}
\affiliation{Dipartimento di Scienze Fisiche, Informatiche e Matematiche, Universit$\grave{a}$ di Modena e Reggio Emilia, I-41125 Modena, Italy}
\affiliation{Centro S3, CNR-Istituto Nanoscienze, I-41125 Modena, Italy}
\author{Marilia J. Caldas}
\affiliation{Instituto de F\'\i{}sica, Universidade de S{\~a}o Paulo,
  Cidade Universit\'aria, 05508-900 S{\~a}o Paulo, Brazil}
\author{Andrea Ferretti}
\affiliation{Centro S3, CNR-Istituto Nanoscienze, I-41125 Modena, Italy}
\author{Daniele Varsano}
\email[corresponding author:]{daniele.varsano@nano.cnr.it}
\affiliation{Centro S3, CNR-Istituto Nanoscienze, I-41125 Modena, Italy}

\date{\today}

\begin{abstract}
Monolayer C$_3$N is an emerging two-dimensional indirect band gap semiconductor with interesting mechanical, thermal, and electronic properties.
In this work we present a description of C$_3$N electronic and dielectric properties, focusing on the so-called momentum-resolved exciton band structure.
Excitation energies and oscillator strengths are computed in order to characterize bright and dark states, and discussed also with respect to the crystal symmetry.
Activation of excitonic states is observed for finite transferred momenta: Indeed,
we find an active indirect exciton at $\sim$ 0.9 eV, significantly lower than the direct optical gap of 1.96 eV, with excitonic binding energies in the range 0.6-0.9 eV for the lowest states. 
As for other 2D materials, we find a quasi-linear  
excitonic dispersion close to $\Gamma$, which however shows a downward convexity related to the indirect band gap of C$_3$N as well as to the dark nature of the involved excitons. \\
\end{abstract}

\maketitle
\section{Introduction}
Since the rise of graphene~\cite{Novoselov666,novoselov_two-dimensional_2005}, two-dimensional (2D) materials attract ever increasing attention in materials science. The reduced dimensionality and high surface-volume ratio affect all the properties of these systems leading to unique features. For example, graphene is well known for its superior mechanical stability~\cite{galashev_mechanical_2014} and electron mobility~\cite{electron_mob_graphene}, relevant for applications in optoelectronics, sensing, mechanical and energy storage technologies~\cite{2d_rev,2d_rev_stretch,cheung16}. However, graphene is a zero band-gap semimetal, which limits its real application in digital electronic devices that usually require a semiconducting character. 
Hence the great effort, from both theoretical and experimental communities, to find post-graphene layered materials with a finite band gap appropriate for the selected applications. The list of candidates is very large~\cite{Mounet_2018_v2} and includes, among  others, transition metal dichalcogenides~\cite{Wang+TMD+2012}, phosphorene~\cite{Liu+phosp+2014}, and hexagonal boron nitride~\cite{Dean+hBN+2010}.

Carbon-nitrides C$_x$N$_y$ (\textit{e.g.} CN, C$_2$N, C$_3$N, or C$_3$N$_4$) are gaining significant attention in recent years~\cite{Adekoya_Qian_Gu_Wen_Li_Ma_Zhang_2020}. These are metal-free carbon-based layered materials with comparable (or even superior) structural, chemical, and electronic properties with respect to graphene. While preserving structural affinities with graphene, they generally display a semiconducting behavior. One of the most studied materials in this class is graphitic-C$_3$N$_4$~\cite{xu_C3N4}, as its strong optical absorption in the visible makes it suitable for solar energy conversion and photocatalysis applications~\cite{naseri_rev_C3N4_photoc}.
\begin{figure}[!t]
  \centering
    \includegraphics[width=0.35\textwidth]{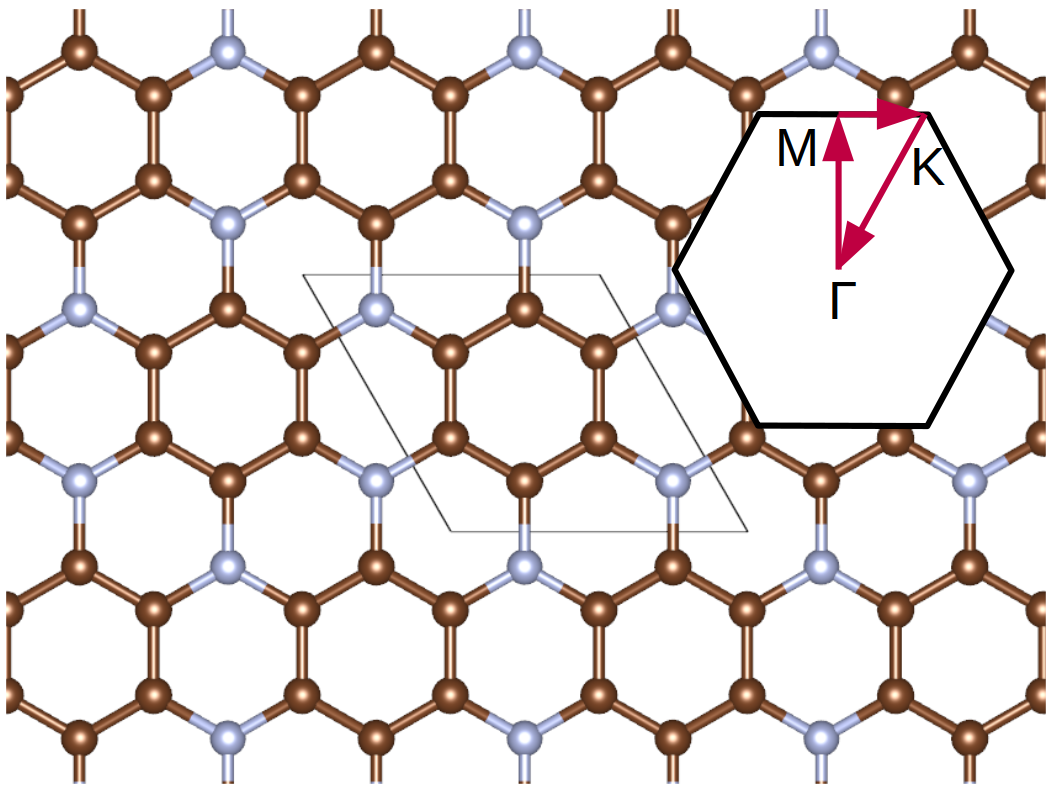}
    \caption{Lattice structure of C$_3$N, with the corresponding unit cell. Brown and grey spheres represent the C and N atoms respectively. Inset: Brillouin zone with the $k$-path considered in this work.
    \label{fig:latt}
    }
\end{figure}
Recently, a new carbon nitride, C$_3$N (structurally analogous to graphene and also known as 2D-polyaniline; see Fig.~\ref{fig:latt}), an indirect gap semiconductor, has been experimentally synthesized using different methods~\cite{Mahmood+c3n+exp,yang+c3n+ferrom}. Since then, several studies were undertaken to characterize it. Theoretical investigations have shown that monolayer C$_3$N displays interesting features: remarkable mechanical properties, similar to graphene~\cite{shi+strain,WANG201822}, high stability at room temperature and high thermal conductivity~\cite{MORTAZAVI201725,Gao2018}. Moreover, it seems a promising material for acting as anode for Li-, Na-, and K-ion  batteries~\cite{xu+battery+exp,guo_c_2019,liu_carbon_2018,bhauriyal_graphene-like_2018}. It was also studied in heterostructures with graphene~\cite{WANG201822}, C$_3$B~\cite{zhang+heteroj}, and g-C$_3$N$_4$~\cite{Wang_2018}, showing interesting properties for optoelectronic and electrochemical energy storage devices. 

Despite these interesting features, an extended characterization of excitations, including excitonic effects, in monolayer C$_3$N is still lacking. 
This knowledge would be relevant when considering C$_3$N for optoelectronic applications like photocatalysis, photodetection, solar cells, or light emitting diodes.
Since the system is a 2D monolayer, the electronic screening is strongly reduced, implying that the enhanced electron-hole interaction typically leads to large excitonic binding energies (EBE) for all the relevant excitons.
In a complementary direction, as
discussed by Qiu \textit{et al.}~\cite{qiu_nonanalyticity_2015} and Cudazzo \textit{et al.}~\cite{cudazzo+exc+2d}, the study of the excitonic dispersion with respect to the transferred momentum $\mathbf{q}$ can give more information on the excitonic features, in particular for low-dimensional systems. Moreover, it may have implications in plasmon/phonon - exciton coupling phenomena and can help in the determination of the exciton propagation along the crystal, relevant for excitation energy transport in the material~\cite{qiu2021signatures}.
In this work we employ many-body perturbation theory methods in the GW approximation~\cite{Hedin_1965,onida_electronic_2002} and the Bethe Salpeter equation (BSE) ~\cite{strinati1988application} to study C$_3$N.
In addition to a full characterization of the electronic and optical properties of this system, the excitonic band structure is computed in order to understand the exciton dispersion  beyond the long wavelength limit. Specifically, momentum-dependent excitonic wavefunctions are also computed and discussed. 
%

\section{Methods}
\label{sec:methods}
%
Ground state electronic structure properties were calculated in the framework of density functional theory (DFT), using a plane wave basis set and pseudopotential approach as implemented in the Quantum ESPRESSO package~\cite{2009qe,2017qe}. We used optimized norm-conserving Vanderbilt (ONCV) pseudopotentials~\cite{oncv1} to compute the relaxed structure and the corresponding Kohn-Sham (KS) single-particle energies and electronic density. 
The KS-DFT exchange-correlation functional was approximated using GGA-PBE~\cite{Perdew-Burke-Ernzerhof1996PRL}.
The KS electronic structure was converged using a plane wave kinetic energy cutoff energy of 60 Ry to represent the wavefunctions, and a Monkhorst-Pack $\mathbf{k}$-point grid of 8$\times$8$\times$1. 
We verified that a supercell with a vacuum of 15~$\text{\AA}$  along the vertical direction was enough to treat the system as 2D, avoiding spurious interactions between layers.

The KS energies are then corrected using many-body perturbation theory methods. In particular, we adopted the GW approximation \cite{onida_electronic_2002} in the single shot G$_0$W$_0$ approach, and then computed the excitonic properties by solving the BSE ~\cite{strinati1988application}.
GW and BSE calculations were 
performed using the Yambo code~\cite{yambo1,yambo2}. 
In the GW approximation, the expectation values of the self-energy are written as:
\begin{equation}
   \Sigma_{n\kk}(\omega) = - \int \frac{d\omega'}{2 \pi i} \, e^{i\omega' 0^+}\, \langle n \kk | \, G(\omega + \omega')W(\omega') \, | n\kk\rangle,
\end{equation}
where $G$ is taken here as the Green's function built on single particle KS orbitals (and their corresponding eigenvalues $\epsilon_{n\mathbf{k}}$) and $W$ is the screened Coulomb interaction  calculated in the random phase approximation (RPA) with the frequency dependence  treated via the Godby-Needs plasmon-pole model~\cite{godby_metal-insulator_1989}.
Quasiparticle energies were then obtained by solving the linearized equation
\begin{equation}
    E_{n\kk}^{\text{qp}} = \epsilon_{n\kk} + Z_{n\kk} \left[\Sigma_{n\kk}(\epsilon_{n\kk})-v^{xc}_{n\kk} \right],
\end{equation}
where $ v^{xc}_{n\kk}$  is the
expectation value of the exchange-correlation potential over the $n\kk$ KS eigenvectors, and $Z_{n\kk}$ is the renormalization factor
\begin{equation}
    Z_{n\kk}=\left[1- \frac{\partial \Sigma_{n\kk}(\omega)}{\partial \omega} \big|_{\omega = \epsilon_{n\kk}} \right]^{-1}.
\end{equation}
In all the calculations we used a truncated Coulomb potential~\cite{rozzi_exact_2006} in order to avoid spurious interaction between repeated cells. 

Converged quasiparticle energies were obtained  using a kinetic energy cutoff of 10 Ry to represent the screening matrix, including 300 empty states in the sums over transitions and adopting a $\mathbf{k}$-point grid of 30$\times$30$\times$1. Convergence on screening matrix cutoff and empty states is reached by means of automated workflows as implemented in the current version of the AiiDA-Yambo plugin~\cite{yambo2} of the AiiDA software~\cite{huber_aiida_2020}, imposing a convergence threshold of 15 meV. More details on the convergence workflow are provided in the \suppinfo.

The quasiparticle energies were then inserted in the BSE. The excitonic Hamiltonian H$^{\text{exc}}$ was built~\cite{onida_electronic_2002,gatti_exciton_2013} considering
a transferred momentum 
$\mathbf{q} = \mathbf{k}_{c}-\mathbf{k}_{v}$
between valence and conduction states,  and reads (in the resonant block):
\begin{equation}
\begin{split}
      \langle vc\kk,\mathbf{q} | H^{\text{exc}} & |v'c'\kk',\mathbf{q} \rangle = \\ & (E_{c',\kk'}^{\text{qp}}-E_{v',\kk'-\mathbf{q}}^{\text{qp}}) \delta_{vc\kk,v'c'\kk'} \\
    & 
    + \langle vc\kk,\mathbf{q} | v - W | v'c'\kk',\mathbf{q} \rangle,
    \label{BSE_q}
\end{split}
\end{equation}
where $|vc\kk,\mathbf{q} \rangle \equiv |c,\kk \rangle \otimes |v,\kk-\mathbf{q} \rangle$, $ |v,\kk-\mathbf{q} \rangle$ and  $|c,\kk\rangle$ being valence and conduction wave-functions, respectively. Concerning the kernel, $W$ and $v$ are the (screened) direct and (bare) exchange e-h interactions.
The BSE can then be recast in the form of a two-particle eigenproblem for the excitonic eigenvalues $E^{S}(\mathbf{q})$ and eigenstates $|S,\mathbf{q} \rangle$:
\begin{equation}
    H^{\text{exc}}(\mathbf{q}) |S,\mathbf{q} \rangle = E^{S}(\mathbf{q}) |S,\mathbf{q} \rangle
    \label{BSE_lambda}
\end{equation}
\begin{equation}
    |S,\mathbf{q} \rangle = \sum_{vc,\kk}  A_{vc\kk}^{S}(\mathbf{q}) |vc\kk,\mathbf{q} \rangle.
\end{equation}
It is worth noticing that the $\mathbf{q}$ index makes the BSE block-diagonal: for different values of the transferred momentum we have different excitonic Hamiltonians $H^{\text{exc}}(\mathbf{q})$. Then, the excitonic band structure is calculated by solving the eigenvalue problem in Eq.~(\ref{BSE_lambda}) for each $\mathbf{q}$ along a path. In this work the BSE was solved for several values of $\mathbf{q}$ along the $\Gamma$-$M$ and $\Gamma$-$K$ directions. 

It is also possible
to define the inverse macroscopic dielectric function (considering only resonant and anti-resonant parts of the H$^{\text{exc}}$) as:
\begin{equation}
\begin{split}
    & \epsilon_M^{-1}(\omega,\mathbf{q}) =  1 +  \frac{2}{V N_q }v(\mathbf{q})\sum_{S}\Omega^{S}(\mathbf{q}) \\
&    \times \left[ \frac{1}{E^{S}(\mathbf{q}) - (\omega+i\eta)} + \frac{1}{E^{S}(\mathbf{q}) + (\omega+i\eta)} \right],
    \label{eps_macro}
\end{split}
\end{equation}
where $V$ is the volume of the unit cell and $N_q$ is the number of points in the Brillouin zone (BZ) sampling.
In the above equation,
$\Omega^{S}(\mathbf{q})$ are the generalized oscillator strengths, defined as:
\begin{equation}
    \Omega^{S}(\mathbf{q}) = \left | \sum_{vc,\kk} A_{vc,\kk}^{S}(\mathbf{q}) \langle v,\kk-\mathbf{q} | e^{-i \mathbf{q}\cdot\rr} | c,\kk \rangle \right |^2 .
    \label{strength}
\end{equation}

Oscillator strengths are used to distinguish between active and inactive excitons (in the $\mathbf{q}=0$ limit, in particular, to identify  those that are dipole active/forbidden). 
Experimentally, excitonic effects are observed via photoabsorption spectroscopy in the optical limit of vanishing $\mathbf{q}$, while non-vanishing momentum excitations are observed \textit{e.g.} in electron energy loss spectroscopy (EELS) or non-resonant inelastic X-ray scattering (NRIXS) experiments, by measuring the loss function L($\omega$,$\mathbf{q}$)~\cite{onida_electronic_2002}. The optical absorption and the loss function are related to the inverse of the macroscopic dielectric matrix $\epsilon_M^{-1}(\omega,\mathbf{q})$ by the relations: 
\begin{eqnarray}
    \text{Abs}(\omega,\mathbf{q} \rightarrow 0) &=& \text{Im}\big[\frac{1}{\epsilon_M^{-1}(\omega,\mathbf{q} \rightarrow 0)} \big],
    \label{eq_abs}
    \\
    \text{L}(\omega,\mathbf{q})  &=&  -\text{Im}\left[\epsilon_M^{-1}(\omega,\mathbf{q}) \right] .
    \label{eq_loss}
\end{eqnarray}
In EELS and NRIXS experiments, the measured quantity is the intensity of the scattered particles resolved with respect to the scattering angle and to the  energy of the outgoing particles (either electrons or photons). This quantity, that is the differential cross-section $\frac{d^2\sigma}{d\Omega dE}$ of the process, can be related to the loss function by the relation~\cite{Sturm+1993+233+242,fossard_angle-resolved_2017}:
\begin{equation}
\label{link_L_cross}
        \frac{d^2\sigma}{d\Omega dE} \sim 
        \left\{ 
            \begin{array}{ c l }
                q^{-2} L(\omega, \textbf{q}) & \quad \text{EELS}, \\ \\
                q^2 L(\omega, \textbf{q}) & \quad \text{NRIXS}.
            \end{array} \right.
\end{equation}
Computationally, in order to converge BSE calculations for C$_3$N, 
we used a $\mathbf{k}$-point mesh of 64$\times$64$\times$1. To numerically represent and solve the eigenproblem of Eq.~(\ref{BSE_q}) we verified that three valence and three conduction bands are enough to obtain converged excitation energies for the first excitons within 10 meV.
\section{Results and discussion}
\label{results}
\begin{figure}
    \centering
    \includegraphics[width=\columnwidth]{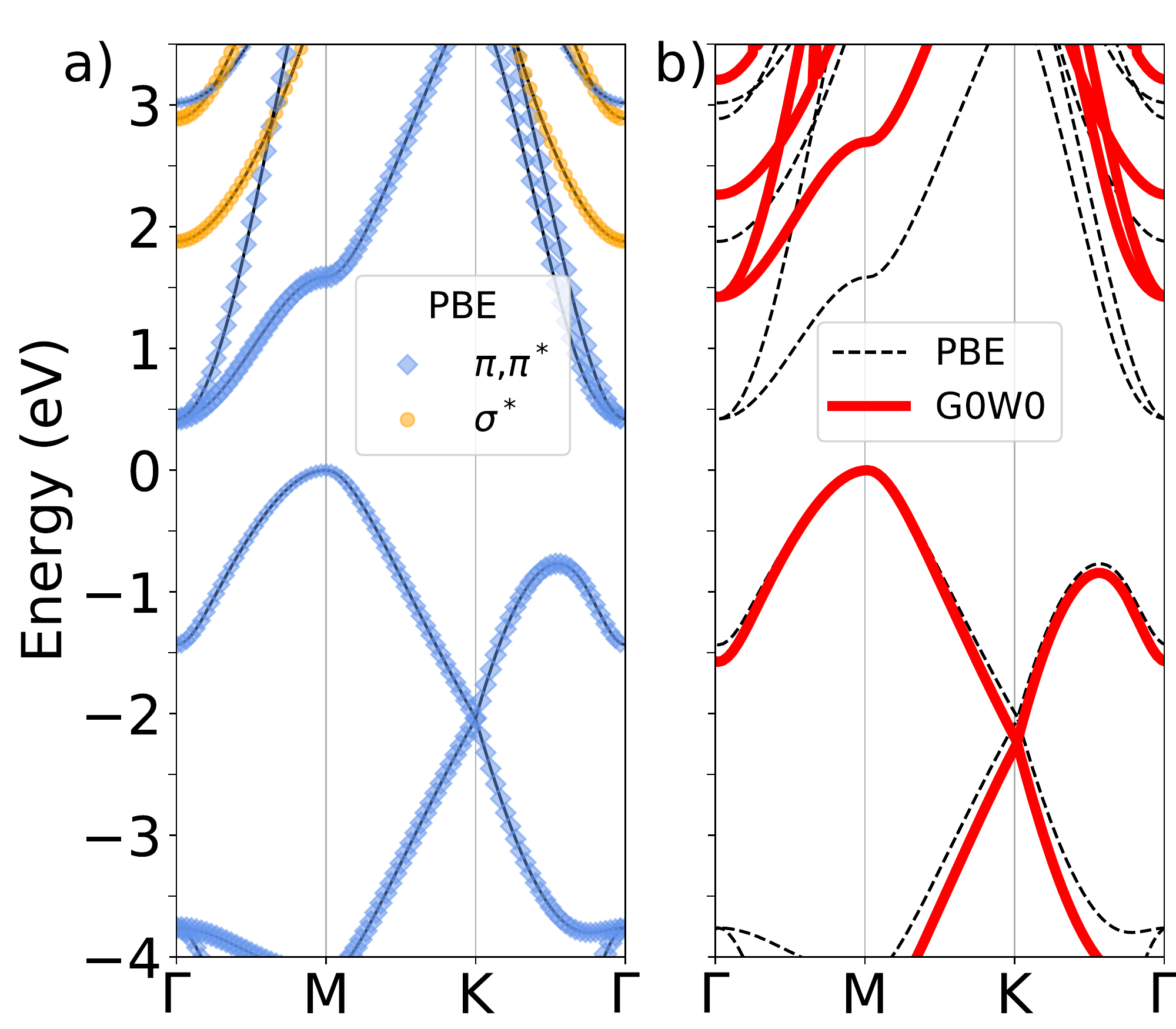}
    \caption{a) Orbital projected PBE band structure. $\pi$ and  $\pi^*$ bands are indicated in blue, $\sigma^*$ as orange dots. 
     b) G$_0$W$_0$ band structure (red); the corresponding PBE bands are reported again as dashed curves for reference. 
     In both panels the zero of the energy scale is set at the top of the valence band for both PBE and G$_0$W$_0$ results.}
    \label{bands}
\end{figure}

%
%
\label{results:qp}
\textbf{Electronic properties.} C$_3$N has a honeycomb lattice, with a unit cell composed of 6 carbon and 2 nitrogen atoms, as shown in Fig.~\ref{fig:latt}. Like graphene, it assumes a planar and continuously bonded geometry, due to the character of the $sp^2$ hybridized orbitals.
For monolayer C$_3$N we find a relaxed lattice constant of 4.857~\AA{} (using GGA-PBE). This result agrees well with previous DFT calculations (4.862~\AA)~\cite{zhou2017} and is slightly larger than the experimental value (4.75~\AA)~\cite{Mahmood+c3n+exp}. 
The KS band structure (Fig.~\ref{bands}a) shows a Dirac cone at the $K$ point. Now, since C$_3$N has two more electrons per unit cell than graphene,  the level-filling goes up and the Dirac cone is no longer at the valence-band top. The calculated band gap is indirect ($M$ $\rightarrow$ $\Gamma$) and amounts to 0.42 eV (at the GGA-PBE level). The direct gaps at $\Gamma$ and $M$ are 1.86 eV and 1.61 eV, respectively.
Figure~\ref{bands}a highlights also the $\sigma$ and $\pi$ character of the bands in the gap region, and we can observe that the top valence bands show $\pi$ symmetry as well as the two lowest conduction bands, whereas $\sigma$-like character is seen in the third and the fourth conduction bands (which start above $\sim$2 eV with respect to the top valence band).

The inclusion of GW corrections provides a gap opening of $\sim$ 1 eV, resulting in an indirect quasiparticle band gap of 1.42 eV (Fig.~\ref{bands}b). The direct gaps at $\Gamma$ and $M$ are also increased to 2.96 and 2.67 eV, respectively, as summarized in Tab.~\ref{table:table_gaps}. The minimum direct gap is located in the middle of the  $\Gamma$-$M$ region of the BZ and has a value of 2.62 eV.
\begin{table}[!t]
    \centering
    \begin{tabular}{c|c|c}
     \hline \hline
      {\bf gaps} & {\bf PBE} (eV) & {\bf G$_0$W$_0$} (eV) \\
      \hline
      M $\rightarrow$ $\Gamma$ & 0.42 & 1.42 \\
      $\Gamma$ $\rightarrow$ $\Gamma$ & 1.86 & 2.96 \\
      M $\rightarrow$ M & 1.61 & 2.67 \\
      \hline \hline
    \end{tabular}
    \caption{Indirect and direct energy gaps at high symmetry points obtained at PBE and G$_0$W$_0$ level.}
    \label{table:table_gaps}
\end{table}
The GW band gap is consistent with the one obtained in a recent work by Wu \textit{et al.}~\cite{wu+BGW}, calculated at the GW level using the Hybertsen-Louie generalized plasmon pole model~\cite{hybertsen_electron_1986}, and as expected is larger than that obtained directly with the HSE06 hybrid functional\cite{krukau2006influence}, by $\sim$ 0.38 eV~\cite{zhou2017}.
The C$_3$N gap is narrower with respect to the two well-known carbon nitrides g-C$_3$N$_4$ ($\sim$4.24 eV for the triazine structure)~\cite{Wei_Jacob_2013} and C$_2$N ($\sim$3.75 eV)~\cite{sun_C2N} single layers. This is consistent with the fact that C$_3$N is continuously bonded, while the two other carbon nitrides have saturated porous structure resulting in electronic confinement. 

\begin{figure}
    \centering
    \includegraphics[width=0.8\columnwidth]{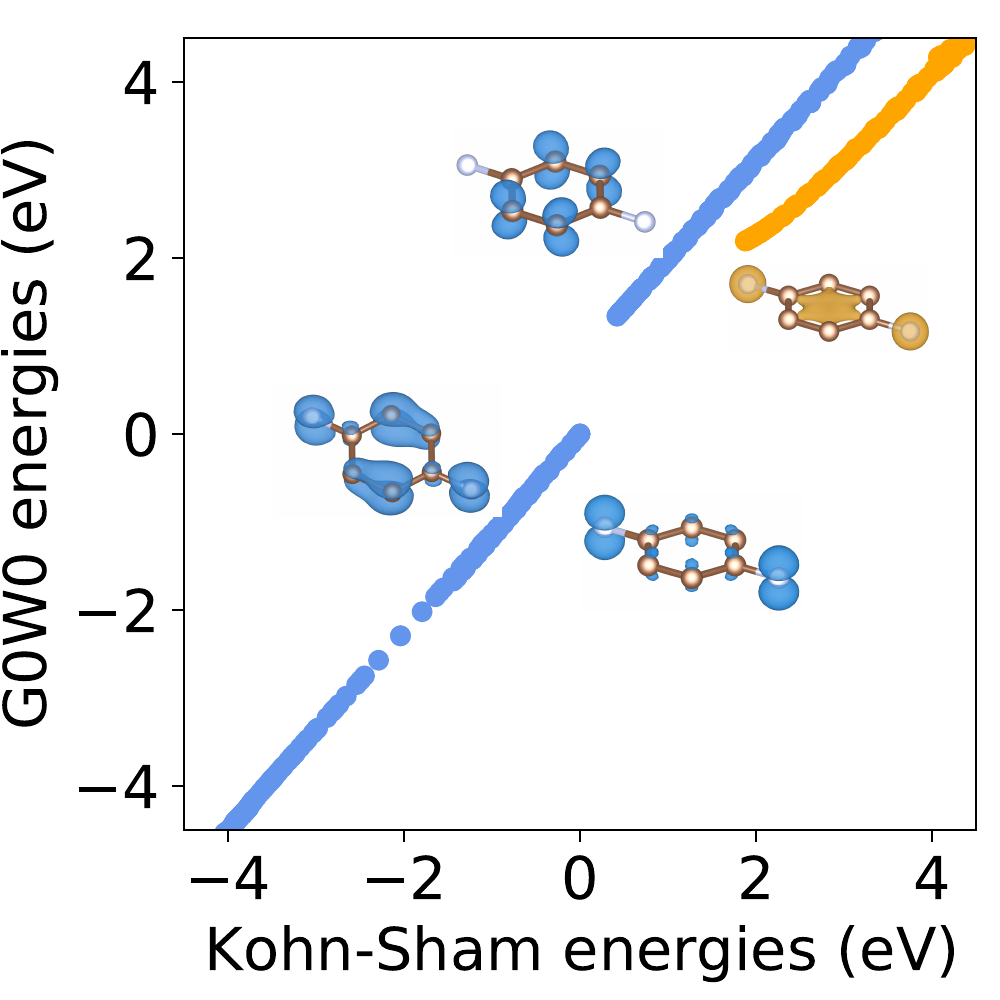}
    \caption{Quasiparticle G$_0$W$_0$ energies vs. Kohn-Sham PBE  eigenvalues. The top of the valence band is chosen as a common zero of the energy. 
    The colours and the corresponding insets indicate the predominant $\pi$, $\pi^*$ (blue) and $\sigma^*$ (orange) character. 
    In the conduction region,  two branches result from the different localization of the  $\pi^*$- and $\sigma^*$-character states leading to different quasiparticle corrections.}
    \label{stretching}
\end{figure}

Along the same BZ path, the C$_3$N band structure calculated at GW level is qualitatively similar to the one calculated using PBE, as shown in Fig.~\ref{bands}b. In fact, the GW gap is still indirect with the valence band maximum located at $M$ and the conduction band minimum at $\Gamma$.
We observe, anyhow,  that the GW correction is not uniform across the bands, as different corrections apply due to wave functions with different character. Orbital-dependent GW corrections are a a well-known feature of GW already discussed in the literature~\cite{yang_excitonic_2007,GNR_Prezzi_2008}.
In particular,  the corrections in the low conduction region are distinguished according to the different localization of the KS wavefunctions, as can be observed from Fig.~\ref{stretching}, where the GW energies are plotted against the respective KS eigenvalues. 

\begin{figure}[!t]
    \centering
    \includegraphics[width=0.8\columnwidth]{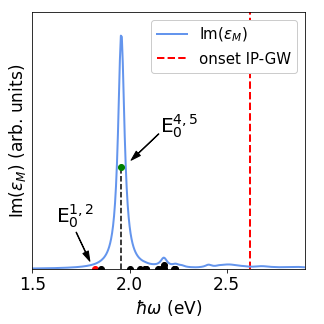}
    \caption{Calculated  Im($\epsilon_M$) in the optical limit ($\mathbf{q} \rightarrow 0 $), in arbitrary units (arb. units). A Lorentzian broadening of 0.02 eV has been included.
    The dots indicate the amplitude of each excitonic state in the BSE solution. The vertical dashed red line represents the onset of the single quasiparticle continuum (minimum direct gap in the GW solution, 2.62 eV).}
    \label{abs}
\end{figure}

A larger correction is observed for the bands with $\pi^*$ (blue dots) character with respect to bands having $\sigma^*$ character (orange dots). Since the valence bands considered here only present a $\pi$ character ($\sigma$-states are deeper in energy), their GW correction shows a more common scissor/stretching behavior. 
We note that this picture allows us to include the GW correction (for instance in the BSE calculations) as a generalized scissor and stretching operator, taking into account different corrections according to the specific localization character of the KS-states. 
\\
%
%
\label{results:optics}
\textbf{Optical properties.} The optical absorption, obtained from the  $\mathbf{q}$ $\rightarrow$ 0 limit of the macroscopic dielectric response, according to Eq.~\eqref{eq_abs}, is shown in Fig.~\ref{abs}.  Excitonic effects were included by solving the BSE equation. All the excitation energies are plotted, together with the independent-particle onset, at E$_{IP-GW}$=2.62 eV, indicated by the vertical red line. The absorption spectrum is dominated by a main peak at 1.96 eV, originated by the doubly-degenerate exciton E$_{0}^{4,5}$. 
At lower energies, three dark excitons are present, the lowest ones (1.82 eV, E$_{0}^{1,2}$) being also doubly-degenerate.

\begin{figure}[!t]
    \centering
    \includegraphics[width=0.9\columnwidth]{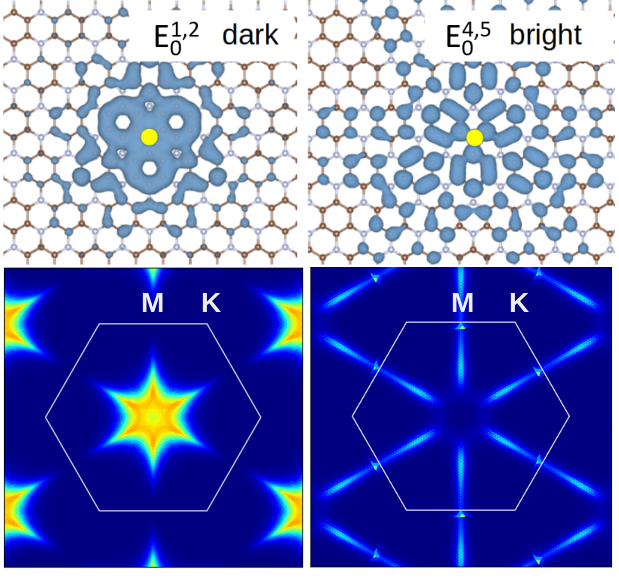}
    \caption{Upper panels: excitonic probability distribution for the dark E$_{0}^{1,2}$ and the bright (1.96 eV, E$_{0}^{4,5}$) excitons at vanishing $\mathbf{q}$. The blue contours represent the probability of finding the electron (hole) when the hole (electron) is fixed on a N atom (yellow circle). Lower panels: distribution of the electron-hole transitions in the BZ for the corresponding excitons of the upper panels.}
    \label{wfc_q0}
\end{figure}

Both bright and dark excitons are strongly bound, displaying EBE of 0.93 eV and 0.66 eV, for E$_{0}^{1,2}$ and E$_{0}^{4,5}$. The EBE are computed considering the difference between the excitation energy of the exciton and the average energy of the involved electron-hole transitions, as shown in the lower panel of Fig.~\ref{wfc_q0}.
Interestingly, the EBE of the lowest bright exciton E$_{0}^{4,5}$ satisfies the 
scaling law EBE $\sim \frac{1}{4}$E$_{IP-GW}$, found valid for 2D single-layered systems.~\cite{lin_scaling_2015,lin_scaling_2017}.

In order to analyze the spatial extension of the main excitons, we plot their excitonic wavefunction, by showing the electron spatial distribution for a fixed hole position. 
The upper panel of Fig.~\ref{wfc_q0} displays the excitonic wavefunction for the lowest (E$_{0}^{1,2}$) and the brightest (E$_{0}^{4,5}$) direct excitons, with the hole placed on a N site. Both excitons show a 3-fold rotational plus a reflection symmetry with respect to the axis in the armchair direction. 
Interestingly, these excitons are localized on benzene rings, avoiding the N atoms. We observe that the dark excitons (lower energy) are strongly localized around the hole, while the active ones (higher energy) are more delocalized. 
%
We can explain the different spatial localization in terms of the single particle transitions that contribute to the excitations, as represented in the lower panels of Fig.~\ref{wfc_q0}. The
first excitons E$_{0}^{1,2}$ are built from electron-hole transitions belonging to a larger region of the Brillouin zone with respect to E$_{0}^{4,5}$, resulting in a larger confinement in real space.
Similarly to the fundamental gap, the optical gap of C$_3$N is also smaller than the ones calculated for g-C$_3$N$_4$ and C$_2$N (by about $\sim$2 eV and $\sim$1 eV~\cite{sun_C2N}, respectively, by considering the first bright excitons), which is consistent with a picture where the presence of saturated pores in each layer of the structure induces wavefunction localization and quantum confinement. 

%
%
\label{results:eels}
\textbf{Indirect excitons.} Besides the optical absorption limit, our understanding of electron-hole excitations in C$_3$N is further enhanced by inspecting the excitonic spectrum over a wider range of transferred momenta $\mathbf{q} = \mathbf{k}_{\text{elec}}-\mathbf{k}_{\text{hole}}$, 
providing us with the exciton dispersion in momentum space.
Indeed, the excitonic dispersion was recently proposed~\cite{cudazzo+exc+2d} as a powerful tool to distinguish
excitons of different character in low-dimensional systems, going beyond the mere evaluation of the EBE.
The solution of the BSE for finite $\mathbf{q}$
then allows us to fully characterize the C$_3$N excitations accessible by means of momentum-resolved electron energy loss spectroscopy or non-resonant inelastic X-ray scattering. 
\begin{figure}[!t]
    \centering
    \includegraphics[width=0.85\columnwidth]{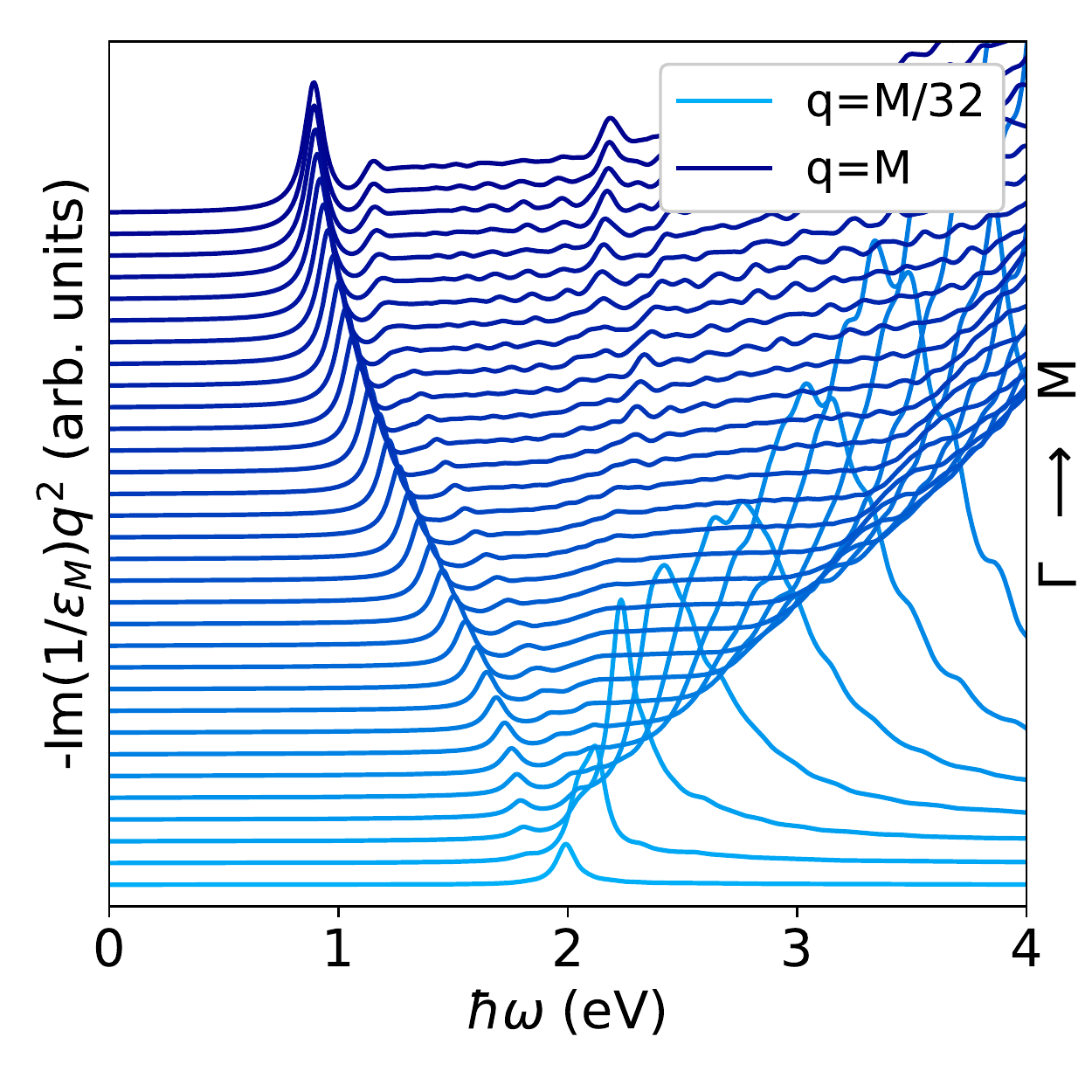}
    \caption{Loss function for different values of the transferred momentum along the direction $\Gamma$-M , starting from the lowest $\mathbf{q}$ = M/32 (light blue), to the highest $\mathbf{q}$ = M (dark blue). Following Cudazzo \textit{et al.}~\cite{cudazzo+exc+2d}, we multiplied all spectra by the corresponding $\mathbf{q}^2$. This is needed in order to enhance higher momentum spectra and physically motivated by the form of the NRIXS cross section (as discussed in Sec.~\ref{sec:methods}).}
    \label{loss_q}
\end{figure}
Figure~\ref{loss_q} shows the loss function calculated for different transferred momenta $\mathbf{q}$ along the $\Gamma-M$ path. To compare spectra at different momentum, we multiplied the loss function by a $q^2$ factor. While making the plot more readable, this factor is also experimentally motivated by the form of the NRIXS cross section, as discussed in the theory section.

We observe that the low-energy peak goes down  in energy for 
$\mathbf{q}$ moving away from $\Gamma$,
as expected for an indirect gap system. 
This indicates the presence of a dispersive excitonic band
reaching a minimum of $\sim$0.8 eV at $\mathbf{q}$ = $M$ in correspondence of the indirect gap. 
\begin{figure*}
    \centering
    \includegraphics[width=1\textwidth]{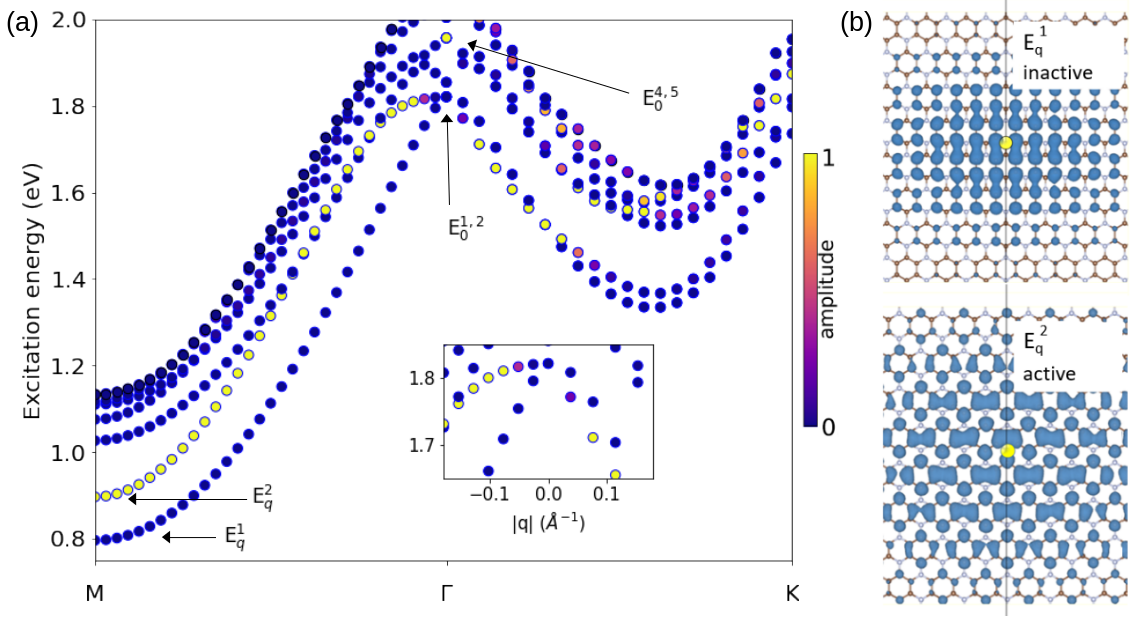}
    \caption{(a) Momentum-resolved exciton band structure for the lowest seven excitons. Excitonic amplitudes are normalized with respect to the brightest exciton among all the bands for each momenta from dark violet to yellow (active exciton). Inset: focus on the region near the q=0 for the two lowest excitonic bands. 
    (b) Exciton probability distribution for E$_{\mathbf{q}}^{1}$ (top) and E$_{\mathbf{q}}^{2}$ (bottom), at $\mathbf{q} = M$. The vertical solid line denotes the symmetry plane perpendicular to the figure and the hole position is indicated in yellow near the center of the figures.
    }
    \label{exc_bands}
\end{figure*}

To complement the information provided by the loss spectra, in Fig.~\ref{exc_bands}a we  present 
the calculated $\mathbf{q}$-resolved exciton band structure. A color map is added to indicate the (generalized) amplitude/strength of each excitonic state, actually offering richer information complementary to the loss spectrum.
The lowest degenerate exciton E$_{0}^{1,2}$ at $\Gamma$ splits into two excitonic bands that we call 
E$_{\mathbf{q}}^{1}$
and E$_{\mathbf{q}}^{2}$. 
For both $\Gamma-M$ and $\Gamma-K$ directions, we observe 
that the lowest excitons present quadratic and almost linear dispersions  in  the  neighborhood  of  the  $\Gamma$  point, as also found for other 2D systems~\cite{qiu_nonanalyticity_2015,sponza+tb+exciton,cudazzo+exc+2d}.
Nevertheless, at variance with what reported in the work of Qiu \textit{et al.}~\cite{qiu_nonanalyticity_2015} for the direct gap MoS$_2$ monolayer, 
here we find a lower band with quasi-linear dependence on $\vert \mathbf{q} \vert $ and an upper band with a clear parabolic dispersion, both with a negative slope and concavity. Such a behavior is due to the indirect nature of the gap, and we note that --at difference with other 2D materials such as MoS$_2$ and phosphorene~\cite{qiu2021signatures}-- in the case of C$_3$N we do not observe a non-analytical dispersion for the lowest bands. This is related to the fact that the $E_0^{1,2}$ excitons are dark, making the exchange contribution to the dispersion negligible. Moreover, the observed down-ward quasi-linear behavior can be seen to be connected to the independent-particle contribution to the excitonic Hamiltonian, as shown in Fig.~S2 in the~\suppinfo.

Moreover, the $\Gamma-M$ direction is of special interest, since it corresponds to $\mathbf{q}$-vector involved in the indirect band gap: the most relevant excitonic transitions occur in that zone, and the decrease of excitation energies going from $\Gamma$ to $M$ reflects the indirect gap of C$_3$N. From Fig.~\ref{exc_bands}a we observe that the lowest exciton (E$_{\mathbf{q}}^{1}$) is inactive, with an energy minimum of 0.8 eV at $\mathbf{q}=M$. For this momentum, the immediately higher-energy  exciton (0.9 eV, E$_{\mathbf{q}}^{2}$) is active.
Concerning the $\Gamma$-$K$ path, we observe a minimum near $K/2$ that corresponds to an inactive state of energy $\sim$1.3 eV.
We find that EBE range from 0.6 eV to 0.93 eV for the lowest excitonic band in the $\Gamma$-$M$ direction (E$_{0}^{1,2}$-E$_{\mathbf{q}}^{1}$).
%
%
%
As we move out of the long-wavelength regime and choose  given directions of the center-of-mass momentum $\mathbf{q}$, we expect some symmetries in the excitonic wavefunctions, like the 3-fold rotational invariance observed in the plots of Fig. \ref{wfc_q0}, to break.
Figure~\ref{exc_bands}b shows the square modulus of exciton wavefunctions (E$_{\mathbf{q}}^{1}$ and E$_{\mathbf{q}}^{2}$) at $\mathbf{q}$=$M$, corresponding to minima in the band structure of Fig.~\ref{exc_bands}a. As before, the hole is placed on a nitrogen atom. The 3-fold rotational symmetry is no longer present, and only the reflection invariance with respect to the armchair direction is maintained.
Both E$_{\mathbf{q}}^{1}$ and E$_{\mathbf{q}}^{2}$ excitons are delocalized along the zig-zag direction.  For the inactive excitons the symmetry axis lies on a nodal plane, where  the probability amplitude of finding electrons vanishes. The spatial delocalization increases with $\mathbf{q}$ and is higher for the indirect active exciton E$_{\mathbf{q}}^{2}$, as happens for the bright excitons in the optical limit case.
The behaviour of the excitonic wavefunctionc of C$_3$N is very similar to that observed in hBN by Sponza \textit{et al.}~\cite{sponza+tb+exciton}. This is because, along the armchair direction, both systems present the same reflection symmetry that characterizes the localization of excitons. 
%

\section*{Conclusions}
%

In this work we have studied the electronic and optical properties of single-layer C$_3$N by using ab initio methods based on DFT and many-body schemes.
In particular, we have computed the 2D band structure of C$_3$N within the GW approximation, revealing an indirect quasi-particle band-gap of 1.42 eV (top of the valence band located at $M$, bottom of the conduction band at $\Gamma$), with direct gaps at $\Gamma$ and $M$ of 2.96 and 2.67 eV, respectively.
The GW corrections to  quasi-particle energies were also discussed in view of the orbital symmetry and localization.

Neutral excitations, as those sampled by optical absorption ($\mathbf{q}=0$) and electron energy loss spectroscopy (finite $\mathbf{q}$), are computed using the Bethe Salpeter equation.
One of the main results of the work comes from the calculation of the full excitonic dispersion of C$_3$N in $\mathbf{q}$-space, giving access to indirect excitons energies and intensities.
These excitons play an important role in phonon-assisted photoluminescence~\cite{schue_bright_2019,Brem_2020}, and they can be efficiently exploited for chemical sensing~\cite{Feierabend_2017}.
Interestingly, while the excitonic dispersion shows a parabolic behavior at $M$, corresponding to the indirect band gap, at $\Gamma$ we observe a degenerate doublet which splits into a parabolic and a quasi-linear dispersing bands, as for other 2D materials. A peculiarity of C$_3$N, connected with its indirect gap as well as with the dark nature of the excitons involved, is that the quasi-linear band has downward convexity, at variance with MoS$_2$ or hBN and the quasi-linear dispersion of the lower excitonic band comes from its independent-particle contribution.
Finally, this work represents a solid starting point to consider the role of excitonic effects of C$_3$N in more specific cases, \textit{e.g.} in describing  exciton-phonon scattering or the spectroscopy of in-plane adsorbates/defects. 

\section{acknowledgement}
This work was partially funded by the EU through the MaX Centre of Excellence for HPC applications (Project No. 824143), and by Regione Emilia Romagna through a FSE PhD fellowship on Big data analysis \& high performance simulations for materials. 
We also thank SUPER (Supercomputing Unified Platform -- Emilia-Romagna) from Emilia-Romagna POR-FESR 2014-2020 regional funds.
MJC acknowledges CNPq, FAPESP and CAPES via the INCT-INEO grant. 
We acknowledge PRACE for awarding us access to the Marconi100 system based in Italy at CINECA.

\bibliography{bib/graphene_2d,
bib/c3n,
bib/codes,
bib/dark,
bib/manual_bib,
bib/finite_q,
bib/mbpt_theory,
bib/HT,
bib/batteries,
bib/misto,
bib/Cnitrides,
bib/photocat,
bib/suppinfo} 

\end{document}


\title{Supplementary Information: Excitonic effects in graphene-like C$_{3}$N}

\author{Miki Bonacci}
\email[corresponding author: ]{miki.bonacci@unimore.it}
\affiliation{Dipartimento di Scienze Fisiche, Informatiche e Matematiche, Universit$\grave{a}$ di Modena e Reggio Emilia, I-41125 Modena, Italy}

\affiliation{Centro S3, CNR-Istituto Nanoscienze, I-41125 Modena, Italy}
\author{Matteo Zanfrognini}
\affiliation{Dipartimento di Scienze Fisiche, Informatiche e Matematiche, Universit$\grave{a}$ di Modena e Reggio Emilia, I-41125 Modena, Italy}
\affiliation{Centro S3, CNR-Istituto Nanoscienze, I-41125 Modena, Italy}
\author{Elisa Molinari}
\affiliation{Dipartimento di Scienze Fisiche, Informatiche e Matematiche, Universit$\grave{a}$ di Modena e Reggio Emilia, I-41125 Modena, Italy}

\affiliation{Centro S3, CNR-Istituto Nanoscienze, I-41125 Modena, Italy}
\author{Alice Ruini}
\affiliation{Dipartimento di Scienze Fisiche, Informatiche e Matematiche, Universit$\grave{a}$ di Modena e Reggio Emilia, I-41125 Modena, Italy}
\affiliation{Centro S3, CNR-Istituto Nanoscienze, I-41125 Modena, Italy}
\author{Marilia J. Caldas}
\affiliation{Instituto de F\'\i{}sica, Universidade de S{\~a}o Paulo,
  Cidade Universit\'aria, 05508-900 S{\~a}o Paulo, Brazil}
\author{Andrea Ferretti}
\affiliation{Centro S3, CNR-Istituto Nanoscienze, I-41125 Modena, Italy}
\author{Daniele Varsano}
\email[corresponding author: ]{daniele.varsano@nano.cnr.it}
\affiliation{Centro S3, CNR-Istituto Nanoscienze, I-41125 Modena, Italy}

\date{\today}

\maketitle


%

\section{Automatic GW convergences}
%

The convergence of the numerical parameters for the GW calculations, with the exception of the $\mathbf{k}$-point grid, was achieved using the automatic workflow implemented in the {\tt aiida-yambo} plugin~\cite{yambo2}. In particular, the automatic workflow was used to converge the kinetic energy cutoff to represent the screening matrix and the number of empty states needed to converge the polarizability and the GW self-energy. These two parameters are strongly interdependent, and cannot be treated separately~\cite{setten}. 

\begin{figure*}
    \centering
    \includegraphics[width=0.7\textwidth]{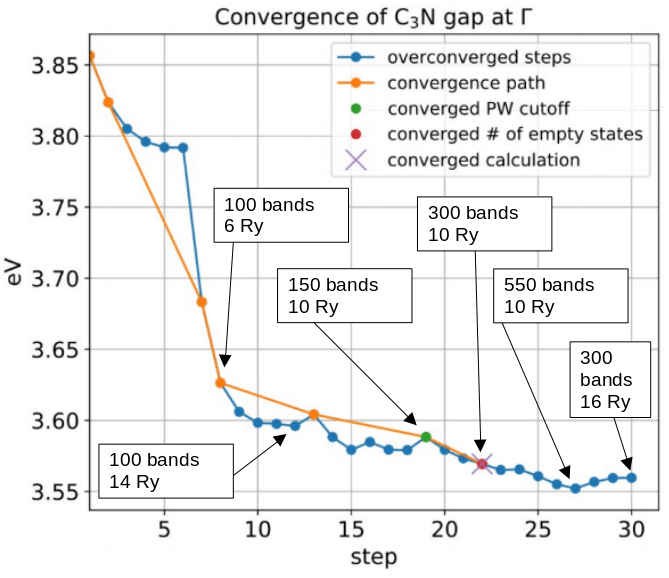}
    \caption{Convergence of the direct gap of C$_3$N at $\Gamma$, with respect to empty states and cutoff on the screening matrix. As shown in the plot, 30 GW runs are sufficient to achieve a convergence of about 30 meV. Further convergence on k-points mesh has been done without the workflow.  
    }
    \label{fig:GW_conv}
\end{figure*}

The workflow used here, as shown in Fig.~\ref{fig:GW_conv}, performs several convergence tests on each parameter, increasing 
one value while keeping fixed the other(s). 
The increment applied in each iteration is provided as input by the user. The convergence threshold is chosen to be $\pm$ 15 meV with respect to the last simulation performed by the workflow, meaning that the final error bar is of 30 meV.

All the calculations are performed taking advantage of other sub-workflows of the plugin, mainly concerning automatic error handling.
%

\section{Independent particle effects in the excitonic band structure}
%

\begin{figure*}
    \centering
    \includegraphics[width=0.6\textwidth]{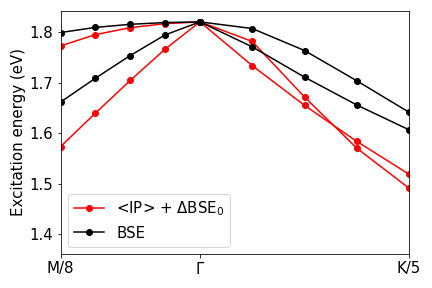}
    \caption{Comparison between independent particle (IP) transition energies (red) and proper excitonic energies (black) for the two lowest excitonic bands. The IP energies are shifted in order to match the excitonic energies at the $\Gamma$ point. We can observe that the negative concavity and the linearity of the lowest excitonic bands are effects already included in the IP dispersion. 
    $\langle \text{IP}\rangle$ and BSE points are aligned at $\mathbf{q}=0$ by including a shift $\Delta\text{BSE}_0$=-0.93 eV.}
    \label{fig:IPvsBSE}
\end{figure*}

As shown in the main text,  close to $\Gamma$ the first two  bands in the excitonic band structure have negative slope and concavity. Moreover, the lowest one has a quasi-linear dependence on $\vert \mathbf{q} \vert $. 
The comparison between the full BSE solution at finite $\mathbf{q}$ close to $\Gamma$ with the average of the independent particle (IP) transition energies, as done in Fig.~\ref{fig:IPvsBSE}, shows that these two peculiar features of the excitonic band structure are inherited from the IP contribution to the Hamiltonian.

Formally, we consider the BSE excitonic Hamiltonian, together with its eigenvectors represented on the transition basis, and separate out its finite-$\mathbf{q}$ contributions, as follows:
\begin{eqnarray}
   H^{\text{exc}}(\mathbf{q}) | S,\mathbf{q} \rangle &=&
    \Omega^{S}(\mathbf{q}) | S,\mathbf{q} \rangle,  \\[5pt]
    |S,\mathbf{q} \rangle &=& \sum_{vc\kk}  A_{vc\kk}^{S}(\mathbf{q}) |vc\kk,\mathbf{q} \rangle, 
\end{eqnarray}
where
\begin{eqnarray}
   H^{\text{exc}}(\mathbf{q}) &=& H^{\text{IP}}(\mathbf{q}) + K_x(\mathbf{q}) + K_d(\mathbf{q}) \\[5pt]
   &=& H^{\text{exc}}(0) + \Delta H^{\text{IP}}(\mathbf{q}) + \Delta K_x(\mathbf{q}) + \Delta K_d(\mathbf{q}) .
   \nonumber
\end{eqnarray}
Then, the average independent particle contribution, $\langle \text{IP}\rangle$, plotted in Fig.~\ref{fig:IPvsBSE} is computed as the expectation value of the diagonal IP term of the BSE excitonic Hamiltonian on the eigenvectors obtained diagonalizing the whole Hamiltonian at each $\mathbf{q}$, namely $\langle \text{IP}\rangle_S = \langle S,\mathbf{q}| H^{\text{IP}}(\mathbf{q}) | S,\mathbf{q} \rangle$.
In practice, these IP energies are evaluated as the weighted energies of all the transitions concurring in the formation of the given exciton as:
\begin{equation}
   \langle \text{IP} \rangle_S = \sum_{cv\mathbf{k}} \vert A^S_{vc\mathbf{k}}(\mathbf{q}) \vert^2 \, (E^{\text{qp}}_{c,\mathbf{k}}-E^{\text{qp}}_{v,\mathbf{k-q}}) .
\end{equation}
Moreover, the $\langle \text{IP}\rangle$ and BSE points in the plot are also aligned at $\mathbf{q}=0$ by a shift $\Delta\text{BSE}_0$ (equal to -0.93 eV, i.e. the EBE of the lowest exciton at $\Gamma$). This amounts to consider the whole BSE Hamiltonian at $\mathbf{q}=0$, while neglecting the $\mathbf{q}$-dependence of $\Delta K_{x,d}$ at finite $\mathbf{q}$, i.e.
\begin{equation}
    \langle \text{IP}\rangle_S + \Delta\text{BSE}_0 = \langle S,\mathbf{q}| H^{\text{exc}}(0) + \Delta H^{\text{IP}}(\mathbf{q}) | S,\mathbf{q} \rangle.
\end{equation}

%

%


\bibliographystyle{aip}
\bibliography{bib/graphene_2d,
bib/c3n,
bib/codes,
bib/dark,
bib/manual_bib,
bib/finite_q,
bib/mbpt_theory,
bib/HT,
bib/batteries,
bib/misto,
bib/Cnitrides,
bib/photocat}